\documentclass{aa}
\usepackage{graphicx}
\begin{document}
\title{The puzzle about the radial cut-off in galactic disks}
\author{C. A. Narayan and C.J. Jog}
\offprints{C. A. Narayan}
\institute{Department of Physics, Indian Institute of Science, Bangalore 560 012, India.\\
\email{chaitra@physics.iisc.ernet.in,cjjog@physics.iisc.ernet.in}}
\date{Received; accepted}

\abstract{
The stellar disk in a spiral galaxy is believed to be truncated physically
because the disk surface brightness is observed to fall faster than that for
an exponential in the outer, faint regions. We review the literature 
associated with this phenomenon and find that a number of recent
observations contradict the truncation picture. Hence we question the very 
existence of a physical outer cut-off in stellar disks. 
We show, in this paper, that the 
observed drop in the surface brightness profiles in fact corresponds to a
negligible decrease in intensity, and that this minor 
change at the faint end appears to be exaggerated on a
log-normal plot.
Since minor deviations from a perfect exponential are common
throughout the disk, we suggest that such a deviation at the faint end 
could easily give rise to the observed sharp drop.
}
\authorrunning{Narayan and Jog}
\titlerunning{Puzzle about the radial cut-off in galactic disks}
\maketitle

\section{Introduction}
One of the important results from the early work on surface photometry 
of several edge-on galaxies by van der Kruit \& Searle (1981) was
the apparent sudden drop
in the surface brightness at a radius of about four disk scale-lengths.
The radius at which this occurs was defined by them to be
$R_{\mathrm{max}}$, or also called $R_{\mathrm{cut-off}}$ in subsequent 
work in the literature.
They argued that this drop represents a physical truncation of
the stellar disk.

This effect was later also observed in many galaxies by a number of
observational groups, both in edge-on galaxies (Jensen \& Thuan 1982, 
Sasaki 1987, Fry
et al. 1999, Barteldrees \& Dettmar 1994, Pohlen et al. 2000a,
de Grijs et al. 2001, Kregel et al. 2002),
and also in face-on galaxies (Pompei \& Natali 1997, Pohlen et al
2002).  For a recent summary
of this topic, see van der Kruit (2001). The truncation
picture proposed by van der Kruit \& Searle (1981) is generally
accepted, though initially it was questioned
(see e.g., van der Kruit 1989, and the discussion thereafter).
Some recent observations have raised doubts about its validity. 
Also, there is no clear physical understanding of the origin
of truncation, although a number of theoretical models have been
proposed for it (Sect. 2.3). 

On looking at the observational data in the literature carefully we  
find that the observed drop in intensity need not imply a  physical
or mass truncation of the disk. The various observational points, questioning
the validity of physical truncation are discussed in Sect. 2.
Some possible solutions are discussed in Sect. 3 and Sect. 4 gives 
a brief summary of our conclusions.

\section{Doubts about outer cut-off} 

\subsection {Questions about the deduction of $R_{\mathrm{max}}$ }

\noindent {\bf 1. Deduction of $R_{\mathrm{max}}$ based on Log-Normal plot:}
The existence of the cut-off radius $R_{\mathrm{max}}$ was first noticed by van der 
Kruit \& Searle (1981)  using the data for the intensity versus
radius plotted on a log-normal plot. 
The use of a log-normal plot is a standard practice in 
the literature and is done for convenience so that several
orders of drop in intensity can be covered with a greater ease.
The logarithm of intensity, measured as surface brightness is 
given in units of magnitude per arc sec$^{-2}$.  We note that
this mode of plotting accentuates any small deviation in the intensity
from an exponential disk at large radii. This is because 
 $d (log I) /d R$ which is 
equal to $(1/I) \: (d I /d R)$ appears to be
very sharp at low values of surface brightness. 
This point is further illustrated next.
 
The observed intensity in a typical face-on spiral galactic disk is
known to obey the following exponential law (Freeman 1970):

$$ I \: = \: I_{\mathrm 0} \: exp \: (- \: R / \: h_{\mathrm R})   \eqno (1) $$

\noindent where $I_{\mathrm 0}$ is the central extrapolated intensity and 
$h_{\mathrm R}$ is the radial disk scalelength.

The intensity profile for an
infinite exponential disk viewed edge-on, $I_{\mathrm{edge-on}} = 
\: (I_{\mathrm 0})_{\mathrm{edge-on}} \: (R/h_{\mathrm R}) \: K_{\mathrm 1} 
(R/h_{\mathrm R})$ 
(van der Kruit \& Searle 1981), where $K_{\mathrm 1}$ is the modifed Bessel function 
of the second kind. For $R >> h_{\mathrm R}$, this reduces to:  

$$ I_{\mathrm{edge-on}} \sim \: (I_{\mathrm 0})_{\mathrm{edge-on}} \: 
(\pi R/2 h_{\mathrm R})^{1/2} exp (-R/h_{\mathrm R}) \eqno (2) $$ 

\noindent Using a constant mass-to-light ratio this gives the 
expression for the edge-on column mass density.
To derive this, the  central disk surface density is taken 
to be $650 \: M_{\odot} pc^{-2}$ as in our Galaxy (from Mera et al. 1998).

Figure 1 shows the column density versus radius for an edge-on infinite 
exponential disk, and
that for a finite disk with a physical truncation at a radius of 20 kpc.
Figure 1a shows these two curves plotted on a standard log-normal plot
while Figure 1b shows the same data on a linear plot. The
difference in the two figures is striking, and clearly shows that
the deduction of $R_{\mathrm{max}}$ would not have been possible if
the data were to be plotted on a linear plot. 
Note that the value of intensity at 20 kpc (for an infinite disk) 
is already very small, $\sim 4 \times 10^{-3}$ times the central value.
Physically, this means that the actual values of the intensity and 
hence the column mass density would not
be substantially different at large radii whether the galactic
disk is taken to be an infinite
exponential disk or whether it has a physical truncation at an
outer radius of $R_{\mathrm{max}}$. 
Hence it is not meaningful
to even define a physical truncation in a galactic disk. 

\bigskip

\noindent {\bf 2. The ratio of $R_{\mathrm{max}} /h_{\mathrm R} $ :} 
The observed profiles for the edge-on galaxies were converted
into the face-on ones using the exponential disk model  
(van der Kruit \& Searle 1981, Sasaki 1987), and the observed  ratio of 
$R_{\mathrm{max}}$/$h_{\mathrm R}$, where $h_{\mathrm R}$ 
is the exponential disk scalelength, was found to be typically $\sim 4$. 
Now,  the central intensity $I_{\mathrm 0}$ (see eq.[1]) is observed to be 
remarkably constant for several 
galaxies (Freeman 1970, van der Kruit 1987, Bosma \& Freeman 1993),
and is equal to 21.65 mag arc sec$^{-2}$.
Thus the Holmberg radius, set by detection limit
to be at 26.5  mag arc sec$^{-2}$, will be equal to
$\sim 4.5$   in units of $h_{\mathrm R}$ as can be obtained from equation (1).
This is too uncomfortably close to the observed range of values
for the ratio of $R_{\mathrm{max}}$/$h_{\mathrm R}$. Hence a simple
explanation for the observed $R_{\mathrm{max}}$ is that the intensity is
too low to be detected beyond this point, so that
$R_{\mathrm{max}}$ does not imply a physical or mass cut-off.

Despite their higher sensitivity, modern observations give 
comparable or even smaller values, in the range of 2-5 for the ratio 
$R_{\mathrm{max}}$/$h_{\mathrm R}$ (Barteldrees \& Dettmar 1994, Pohlen et al. 2000b, 
and de Grijs et al. 2001).
This puzzling point has been commented upon by Barteldrees \& Dettmar (1994).
Thus, the above two points show that contrary to the general belief, 
$R_{\mathrm{max}}$ does not appear to be a 
fundamental parameter of the disk (see Sect. 2.3).

\subsection {Observational evidence against truncation }

\noindent {\bf 1. Dependence of $R_{\mathrm{max}}$ on sensitivity:}  
The size as inferred by the last detectable point or the last
iso-intensity contour of galactic disks is seen to increase when 
observed with a greater sensitivity telescope (Bosma \& Freeman 1993).
This was shown to be true for nearly half of the sample of 222 galaxies 
seen in both the Palomar Sky Survey (with a limiting magnitude, 
$\mu_{\mathrm{lim}} = 24.6 \:$ mag arc sec$^{-2}$) 
and the SRC-J survey ($\mu_{\mathrm{lim}} = 25.6 \:$ mag arc sec$^{-2}$).
The fact that the observed size of a galaxy or equivalently the value of 
$R_{\mathrm{max}}$ increases for a fainter detection limit
 implies that the outer limit
is an artefact of the detection and does not indicate a 
genuine physical cut-off in the disk.
If the truncation were truly a physical one, the size should not vary with
the sensitivity.

\bigskip

\noindent {\bf 2. $R_{\mathrm{max}}$ for the Galaxy:}  
An outer cut-off radius of 15($\pm$ 2) kpc has been deduced for the Galaxy
by modeling the DENIS star-count data by Ruphy et al. (1996). In contrast,
the more recent work by Lopez-Corredoira et al. (2002)
based on the modeling of 2MASS star-counts data in 820 regions
has shown that there is no abrupt cut-off in the stellar disk at
least to within a radius of 15 kpc studied by them. 

This last conclusion agrees with our theoretical result (Narayan \& Jog 2003) 
based on a different method. In that paper, we have argued that the 
variation in the scaleheight with radius for the atomic hydrogen 
data allows us to convincingly rule out a physical cut-off in our 
Galaxy upto 20 kpc.
Since the Galactic disk scalelength, h$_{\mathrm R}, is \sim 3.2$ kpc 
(Mera et al. 1998), this means that the
stellar disk shows no signs of a cut-off till about 6 disk scalelengths.

\bigskip

\noindent {\bf 3. Face-on galaxies:}
The integrated column density or surface brightness is slightly larger
(by a factor of $R^{1/2}$)  for an edge-on galaxy than for a face-on  galaxy 
- compare equations (1) and (2). However, it is this
small difference which allows an easier detection of the faint, outer
regions.
This is the reason why the edge-on galaxies were first chosen 
for photometric studies by van der Kruit \& Searle (1981).
Theoretically, following the same steps as in Sect. 2.1, we
find that a truncation is also seen in a
face-on galaxy. Some face-on galaxies (Shostak \& 
van der Kruit 1984, Pompei \& Natali 1997, Pohlen et al. 2002)
do show evidence for a radial cut-off {\it well beyond} four disk 
scale-lengths.
 Hence, it is not clear why a much larger sample 
of 86 face-on spiral galaxies studied by de Jong \& van der Kruit (1994) 
do not show a cut-off.

\bigskip

\noindent {\bf 4. Rate of decrease of surface brightness:}
The drop in the surface brightness leading to the determination
of R$_{\mathrm{max}}$ does not seem to show a uniform 
well-defined behaviour in all the cases where it is observed. The
rate of fall in intensity is high ($> 1 \:$ mag arc sec$^{-2}$ kpc$^{-1}$) in 
some galaxies (Florido et al. 2001, Barteldrees \& Dettmar 1994) 
which thus show a sharp
cut-off, whereas the rate of fall is rather low ($< 1 \:$
mag arc sec$^{-2}$ kpc$^{-1}$) in most other galaxies (van der Kruit \& 
Searle 1981, Pohlen et al. 2002, Kregel et al. 2002). 
The drop in surface brightness in the outer radii in galaxies has over
the years been also modeled as a double exponential (e.g. Pohlen et
al. 2002). 
There has been no attempt so far to relate this variation to
either the disk formation process or any theoretical truncation 
model. 

\subsection {Theoretical models for $R_{\mathrm{max}}$}  
The origin of the physical truncation in a disk is not clearly
understood yet. Some ideas proposed so far include: critical density
for star formation (e.g., Kennicutt 1989, van den Bosch 2001), the maximum 
angular momentum per unit mass in a protogalaxy (van der Kruit 1989), 
incomplete disk 
formation, expulsion of stars controlled by magnetic field (Battaner
et al. 2002), etc.
The ratio $R_{\mathrm{max}}/ h_{\mathrm R}$ forms an important parameter in these models. 
However, viscous
evolution, which is the currently accepted scenario for the disk
formation, does not lead to the existence of $R_{\mathrm{max}}$ 
{\it naturally} (Saiz et al. 2001).   
But, unfortunately, despite its uncertain status, mass truncation has been 
artificially introduced by both theorists (Casertano 1983, van den
Bosch 2001, Bell 2002) and observers (Bottema 1996, Fry et al. 1999,
Pohlen et al. 2000a, de Grijs et al. 2001)
alike to match their models with obervations.
The physical truncation has even been applied to explain the observed decrease
in the rotation curve near the $R_{\mathrm{max}}$ region for two galaxies 
(Casertano 1983, Bottema 1996). 

\section {\bf Alternative solutions and implications}

As shown in Sect. 2.1, what is seen as a major deviation in a
surface brightness profile far from the centre is in fact a very
small difference in the intensity profile.
In this section we discuss two possibilities other than truncation
that can explain this drop. 

\bigskip

\noindent {\bf (a). Background sky subtraction :} First, we consider the simplest possible reason, 
namely, whether the sharp fall-off could result during the
process of data reduction- as say by an overestimate of sky
brightness during background sky subtraction (see e.g., Binney
\& Merrifield 1998).  We illustrate this point quantitatively in
Figure 2 to show how $R_{\mathrm{max}}$ can arise artificially during the
sky subtraction process.
 In Figure 2, we plot the column density versus radius for an
infinite exponential disk. The column density can be converted to 
surface brightness using standard techniques. The dashed line is the
result of subtraction of correctly estimated background and the solid
line results when a slightly over-estimated background (by 0.5 \%)
has been subtracted.  Although the overestimated sky background is 
subtracted from raw data at all points, the reduction is more important 
in the outer, fainter regions.
The resulting distribution is strikingly similar
to the theoretical profile with physical truncation (Fig. 1a) and to
an observed profile from which physical truncation was deduced
by van der Kruit \& Searle (1981). Thus the observed drop can
be easily explained as arising due to the inaccurate subtraction of the sky
brightness, rather than due to  physical truncation. A smaller
overestimate of 0.1 \% would shift the location of drop to a
larger radius of $\sim$ 30 kpc, while conversely a larger overestimate
of 2.5 \% is needed to give a cut-off at 20 kpc  seen in Figure 1a.

We find that the observers are aware of this 
possibility and take utmost care in the estimation of the true 
background - see Barteldrees \& Dettmar (1994) for a detailed 
discussion on this.
In the past, some arguments against the role of sky subtraction
in causing a drop have been raised.
First, in a few galaxies  the deviation from the exponential 
 happens much before the sky brightness errors can affect the profile 
 (van der Kruit \& Searle 1981, Barteldrees \& Dettmar 1994).
However, this need not be true for all galaxies.
 Second, it
has been argued that if $R_{\mathrm{max}}$ arises due to errors in sky 
subtraction then we should see a similar cut-off in other luminosity 
distributions like the vertical profiles of disks, which is not seen 
(de Grijs et al 2001, Barteldrees \& Dettmar 1994). 
Note, however, that the intensity 
falls much faster along the $z$-axis 
than along $R$-axis for a thin galactic disk. For example, for the galaxy 
ESO 187-G08 (Barteldrees \& Dettmar 1994), the rate of
decrease is 2.63 mag arc sec$^{-2}$ kpc$^{-1}$ along $z$-axis as opposed 
to a change of 0.35 mag arc sec$^{-2}$ kpc$^{-1}$ radially, that is,  
about 7 times higher. This natural rapid fall in 
intensity can easily hide an artificial cut-off along $z$-axis, if any.
Thus, we have argued that the role of sky subtraction in causing truncation 
cannot yet be ruled out.

We would like to point out that the surface brightness profiles for the 
low-luminosity 
elliptical galaxies  also show curves going downwards away from  
the typical $R^{1/4}$ de Vaucouleurs law at all radii (Binney \& Merrifield 
1998), also see  (Binggeli et al. 1984, Prugniel et al.
1993). It is well-known that even some
giant elliptical galaxies such as NGC 4472 are better fitted by
a truncated King's  profile rather than by a de Vaucouleurs profile
(Mihalas \& Binney 1981).
To our knowledge, the topic of radial cut-off in elliptical
galaxies has not been studied systematically in 
the literature, unlike the truncation in spiral galaxies.

\bigskip

\noindent {\bf (b). Intrinsic variation in disk intensity :} The second possibility for the observed cut-off comes 
from the intrinsic nature of 
the distribution of light in the disk itself. 
It is well known that galactic disks are not perfectly exponential
in nature as seen in many studies involving large samples, and minor 
deviations from the exponential (of the order of 0.1-0.5 mag
arc sec$^{-2}$) are commonly seen (e.g., Boroson 1981,  Kodaira
et al.  1986). Departures from the exponential could be due to recent 
star formation 
or oval distortions, bars, spiral arms etc falling along the line of 
sight (Boroson 1981).
In the outer regions, the more likely reason is the presence of 
clumpy star formation,
spiral arms, or ring-like features,  beyond which there could be an
abrupt drop in surface brightness. We point out that this aspect,
though probably
as important as the uncertainty in the sky brightness
subtraction as a possible cause of the spurious cut-off,
seems to have been ignored in the literature.

This approach also explains the following related observation.
In face-on galaxies, the surface brightness profiles are generally 
azimuthally averaged (de Jong \& van der Kruit 1994) thus minimising the
deviations, which could  explain why $R_{\mathrm{max}}$ is not commonly seen in
face-on galaxies.

\section{Conclusions} 
The aim of this paper is to point out that there is no
conclusive evidence for the existence of a physical cut-off in
the stellar disk in the outer galactic disks. 
Some recent observations rule out the abrupt mass truncation
both in our Galaxy as well as in external galaxies.
Instead, we argue that the
observed drop in surface brightness appears more significant
than it is because of the use of a log-normal plot used
routinely in the literature. 
We discuss two probable causes: a small error (of $ < 1$ \%) in sky
brightness subtraction, or the genuine wiggly nature of light
distribution in galactic disks, which
can give rise to a spurious drop in surface
brightness. This topic deserves further study. This paper also shows that 
the various theoretical models proposed so far to explain the origin of 
truncation are not necessary since the
existence of the phenomenon itself is in doubt. Thus we have
here  the proverbial "emperor's clothes" problem on hand.

\bigskip

\noindent {\it Acknowledgements:} We would like to thank the 
anonymous referee, 
and F. Schweizer, for critical comments on the manuscript. We would also 
like to thank T. Prabhu, M. Gopinathan, and A. Subramanian for useful 
discussions.

\bigskip

\noindent {\bf {References}}

\noindent Barteldrees, A., \& Dettmar, R.-J. 1994, A\&AS, 103, 475

\noindent Battaner, E., Florido, E., \& Jimenez-Vicente, J. 2002, A\&A, 388, 
 213

\noindent Bell, E.F. 2002, ApJ, 581, 1013

\noindent Binggeli, B., Sandage, A., \& Tarenghi, M. 1984, AJ, 89, 64

\noindent Binney, J. \& Merrifield, M. 1998, Galactic Astronomy
 (Princeton: Princeton University Press)

\noindent Boroson, T. 1981, ApJS, 46, 177

\noindent Bosma, A., \& Freeman, K. C. 1993, AJ, 106, 1394

\noindent Bottema, R. 1996, A\&A, 306, 345

\noindent Casertano, S. 1983, MNRAS, 203, 735

\noindent de Grijs, R., Kregel, M., \& Wesson, K.H. 2001, MNRAS, 324, 1074

\noindent de Jong, R.S., \& van der Kruit, P.C. 1994, A\&AS, 106, 451

\noindent Freeman, K.C. 1970, ApJ, 160, 811

\noindent Florido, E., et al. 2001, A\&A 378, 82

\noindent Fry, A.M., Morrison, H.L., Harding, P., Boroson, T.A. 1999, AJ, 
118, 1209

\noindent Jensen,  E.B., \& Thuan, T.X. 1982, ApJS, 50, 421

\noindent Kennicutt, R.C. 1989, ApJ, 344, 685

\noindent Kodaira, K., Watanabe, M., \& Okamura, S. 1986, ApJS, 62, 703

\noindent Kregel, M., van der Kruit, P.C. \& de Grijs, R. 2002, 
 MNRAS, 334, 646

\noindent Lopez-Corredoira, M., Cabrera-Lavers, A., Garzon, F., \&
  Hammersley, P.L. 2002, A\&A, 394, 883

\noindent Mera, D., Chabrier, G., \& Schaeffer, R. 1998, A\&A, 330, 953

\noindent Mihalas, D. \& Binney, J. 1981 Galactic Astronomy 
  (San Francisco: Freeman)

\noindent Narayan, C.A., \& Jog, C.J. 2003, in preparation

\noindent Pohlen, M., Dettmar, R.-J., Lutticke,R., \& Schwarzkopf, U. 2000a, 
A\&AS, 144, 405

\noindent Pohlen, M., Dettmar, R.-J., \& Lutticke,R. 2000b, A\&A, 357, L1

\noindent Pohlen, M., Dettmar, R.-J., Lutticke,R., \& Aronica,
G. 2002, A\&A, 392, 807

\noindent Pompei, E., \& Natali, G. 1997, A\&AS, 124, 129

\noindent Prugniel, Ph., Bica, E., Klotz, A., \& Alloin, D. 1993, A\&AS, 98, 229

\noindent Ruphy, S., et al. 1996, A\&A, 313, L21

\noindent Saiz, A., Dominguez-Tenreiro, R., Tissera, P.B., Courteau, S. 
2001, MNRAS, 325, 119

\noindent Sasaki, T., 1987, PASJ, 39, 849

\noindent Shostak, G. S., \& van der Kruit, P. C. 1984, A\&A, 132, 20

\noindent van den Bosch, F.C. 2001, MNRAS, 327, 1334
 
\noindent van der Kruit, P.C. 1987, A\&A, 173, 59 

\noindent van der Kruit, P.C. 1989, in The World of Galaxies,
eds. H.G. Corwin, Jr. \& L. Bottinelli (New York: Springer-Verlag), 
256

\noindent van der Kruit, P.C., \& Searle, L. 1981, A\&A, 95, 105

\noindent van der Kruit, P.C., 2001, in Galaxy disks and disk galaxies,
ASP Conf Ser 230, eds. J.G. Funes, S.J. \&  E.M. Corsini (
San Francisco: ASP), 119

\bigskip

\noindent {\bf Figure 1a.} A plot of edge-on column mass density 
(in $M_{\odot} pc^{-2}$ ) versus radius (in kpc) for an infinite exponential 
disk (dashed line), and for an exponential disk truncated at a radius 
of 20 kpc (solid line), both plotted on a log-normal plot. 
The deviation of the latter from an infinite disk in regions of
low surface density case is accentuated in this mode of plotting. 
It is such deviation, and the subsequent sharp drop, that is 
generally taken to indicate the existence of an outer cut-off in the disk.

\bigskip
\noindent {\bf Figure 1b.} The same data as in Fig. 1a, except here
the curves are plotted on a linear plot. The two curves nearly overlap 
and the difference between
them is not discernible. Hence, from this plot the existence
of an outer cut-off cannot be deduced.

\bigskip
\noindent {\bf Figure 2.} The column density is plotted against the 
radius for an infinite exponential disk seen edge-on. The dashed line is the 
result  of subtraction of correctly estimated sky background while the solid 
line is the result of a slight (0.5 \%) overestimation. The slight 
overestimate could easily be misinterpreted as a physical 
cut-off (compare with Fig. 1a).  Thus, this figure illustrates 
the importance of the correct sky background estimation in bringing out the
true profile of a galaxy.

\newpage
\begin{figure}
{\resizebox{9cm}{9cm}{\includegraphics{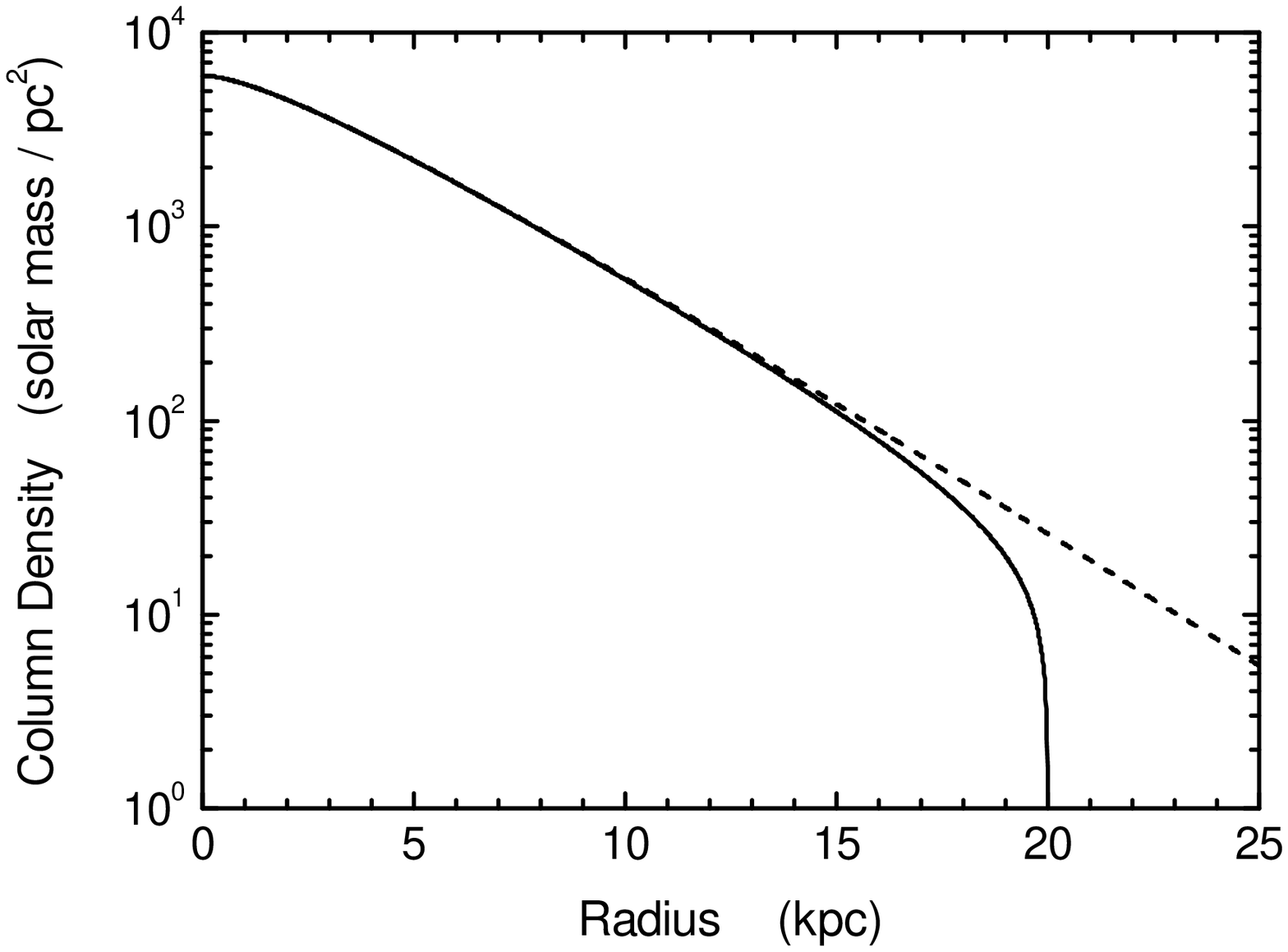}}}
\begin{center}
{\huge Figure 1a.}
\end{center}
\end{figure}

\begin{figure}
{\resizebox{9cm}{9cm}{\includegraphics{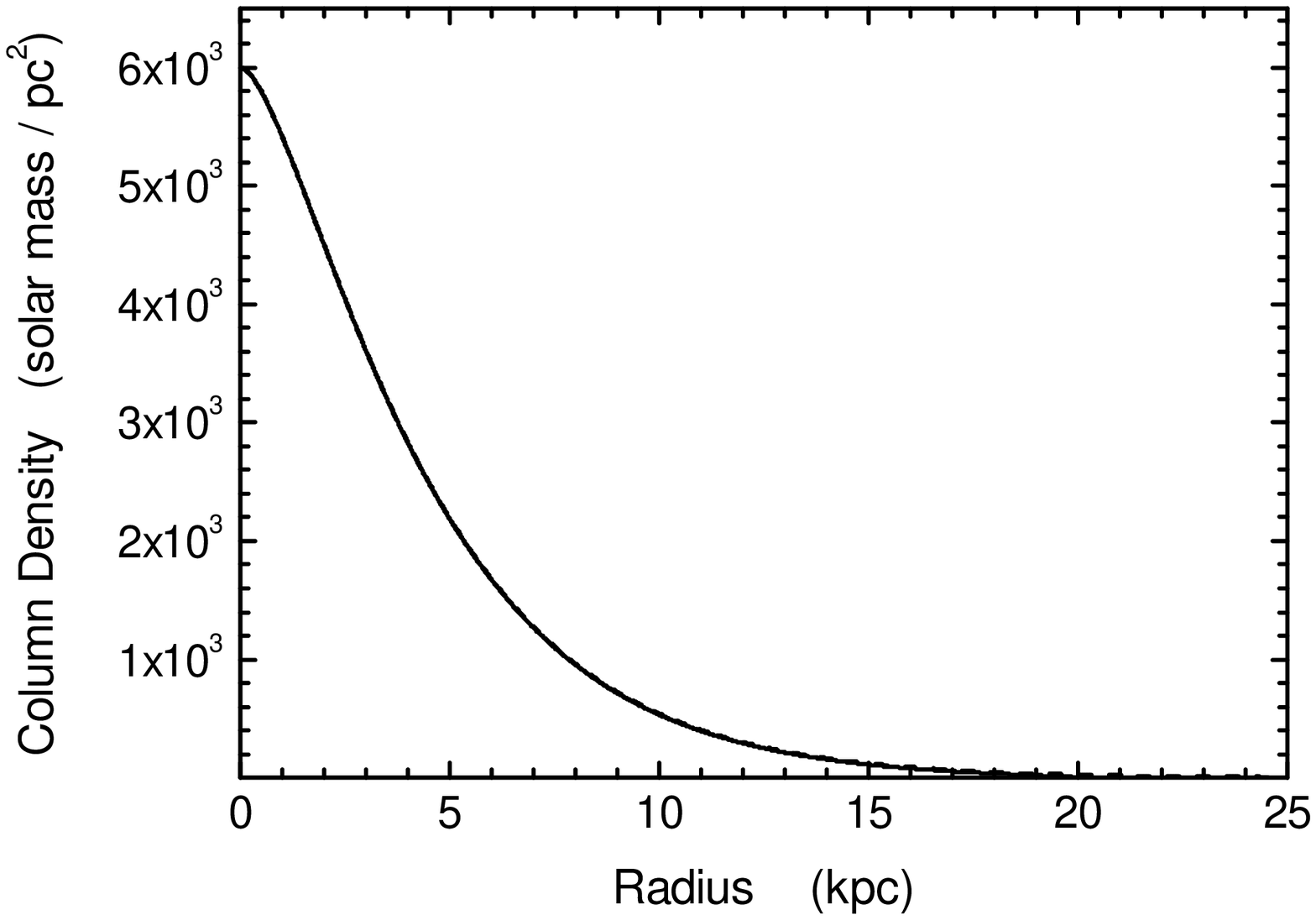}}}
\begin{center}
{\huge Figure 1b.}
\end{center}
\end{figure}

\begin{figure}
{\resizebox{9cm}{9cm}{\includegraphics{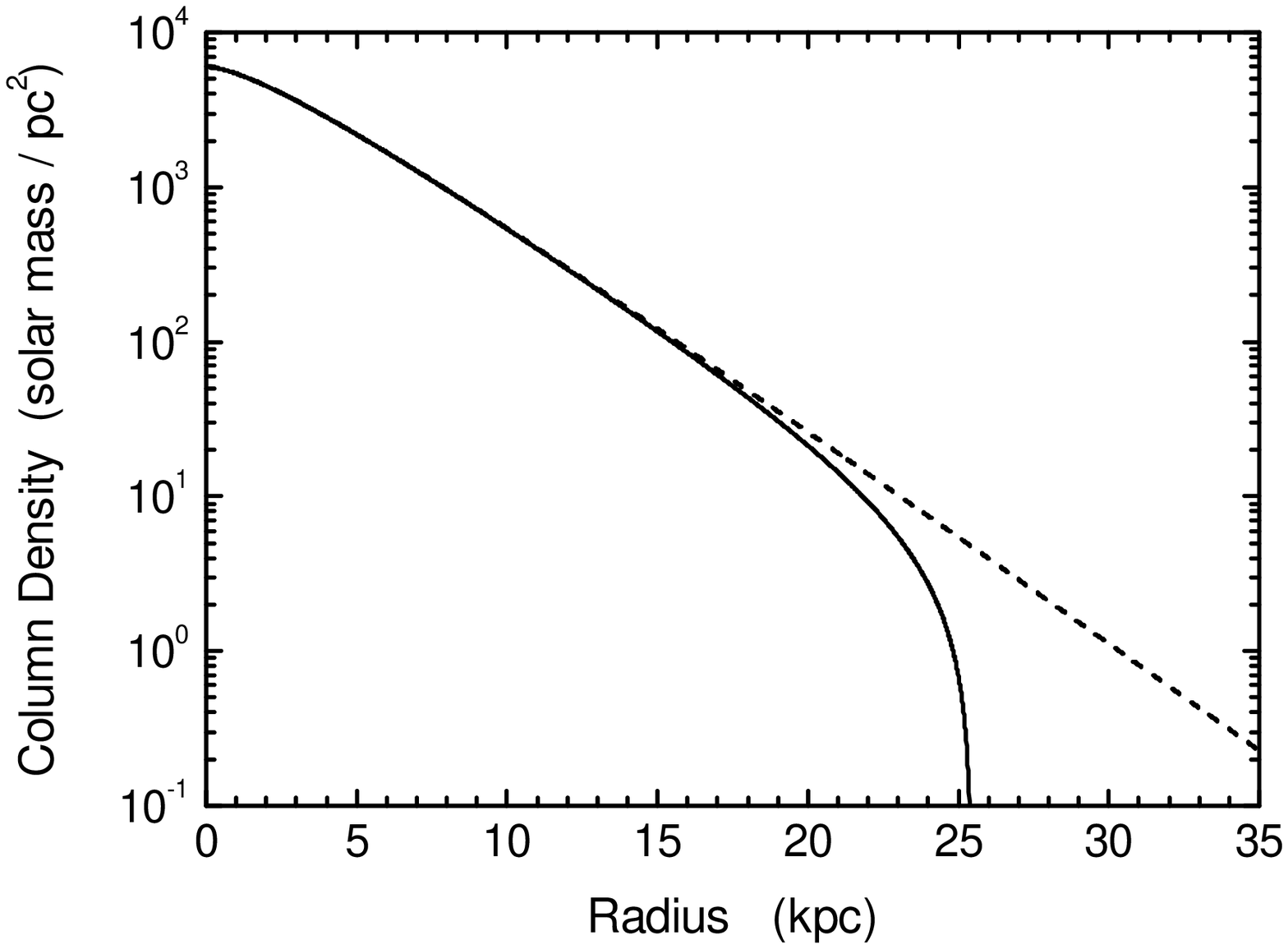}}}
\begin{center}
{\huge Figure 2.}
\end{center}
\end{figure}
\end{document}